\documentclass{aa}
\usepackage{graphicx} %
\usepackage{amsmath} %
\usepackage{amssymb} %
\usepackage{bm} %
\usepackage{multirow}
\usepackage[colorlinks=true,allcolors=blue,urlcolor=blue]{hyperref}
\usepackage{txfonts}

\newcommand{\fig}[1]{\text{Fig.~\ref{#1}}}

\newcommand{\teff}{\ensuremath{T_{\mathrm{eff}}}}
\newcommand{\logg}{\ensuremath{\log{g}}}
\newcommand{\feh}{\ensuremath{\mathrm{[Fe/H]}}}
\newcommand{\vmic}{\ensuremath{{\xi}}}
\newcommand{\xfe}[1]{\ensuremath{\mathrm{[#1/Fe]}}}
\newcommand{\ax}[1]{\ensuremath{A(\mathrm{#1})}}
\newcommand{\xh}[1]{\ensuremath{\mathrm{[#1/H]}}}
\newcommand{\objname}{D25\_6334}
\newcommand{\gaians}{E24\_6328}
\newcommand{\nscanone}{E24\_1350}
\newcommand{\nscanfive}{E24\_5136}
\newcommand{\kms}{\ensuremath{\mathrm{km\,s^{-1}}}}

\defcitealias{El-Badry2024a}{E24}
\defcitealias{Li2018a}{L18}
\defcitealias{Dodd2024a}{D25}

\begin{document}

\title{Unevolved Li-rich stars at low metallicity: a possible formation pathway through novae} 
\author{
    Tadafumi Matsuno\inst{\ref{ari}} \and 
    Alex Kemp\inst{\ref{IvS}} \and 
    Ataru Tanikawa\inst{\ref{Fukui}} \and 
    Cheyanne E. Shariat{\inst{\ref{caltech},\ref{ucla}}} \and 
    Kareem J. El-Badry{\inst{\ref{caltech}}} \and 
    Emma Dodd\inst{\ref{kapteyn},\ref{durham}} \and 
    Amina Helmi\inst{\ref{kapteyn}} \and
    Andreas J. Koch-Hansen\inst{\ref{ari}} \and 
    Natsuko Yamaguchi\inst{\ref{caltech}} \and
    Hongliang Yan\inst{\ref{naoc},\ref{cas}}
}
\institute{
\label{ari}Astronomisches Rechen-Institut, Zentrum f\"ur Astronomie der Universit\"at Heidelberg, M\"onchhofstra{\ss}e 12-14, 69120 Heidelberg, Germany \email{matsuno@uni-heideberg.de} \and 
\label{IvS} Institute of Astronomy (IvS), KU Leuven, Celestijnenlaan 200D, 3001, Leuven, Belgium \and
\label{Fukui} Center for Information Science, Fukui Prefectural University, 4-1-1 Matsuoka Kenjojima, Eiheiji-cho, Fukui 910-1195, Japan \and
\label{caltech} Department of Astronomy, California Institute of Technology, 1200 East California Boulevard, Pasadena, CA 91125, USA \and
\label{ucla} Department of Physics and Astronomy, University of California, Los Angeles, CA 90095, USA \and
\label{kapteyn} Kapteyn Astronomical Institute, University of Groningen, Landleven 12, 9747 AD Groningen, The Netherlands\and
\label{naoc} CAS Key Laboratory of Optical Astronomy, National Astronomical Observatories, Beijing 100101, China \and
\label{cas} School of Astronomy and Space Science, University of Chinese Academy of Sciences, Beijing 100049, China
\label{durham} Institute for Computational Cosmology, Department of Physics, Durham University, South Road, Durham DH1 3LE, UK
}
\titlerunning{Unevolved Li-rich stars}

\abstract{
A small fraction of low-mass stars have been found to have anomalously high Li abundances.
Although it has been suggested that mixing during the red giant branch phase can lead to Li production, this method of intrinsic Li production cannot explain Li-rich stars that have not yet undergone the first dredge-up. 
To obtain clues about the origin of such stars, we present a detailed chemical abundance analysis of four unevolved Li-rich stars with $-2.1 < \feh < -1.3$ and $2.9<\ax{Li}<3.6$, $0.7-1.4$ dex higher Li abundance than typical unevolved metla-poor stars.
One of the stars, Gaia DR3 6334970766103389824 (\objname), was serendipitously found in the stellar stream ED-3, and the other three stars have been reported to have massive ($M\gtrsim 1.3\,\mathrm{M_\odot}$) non-luminous companions. 
We show that three of the four stars exhibit abundance patterns similar to those of known unevolved Li-rich stars, namely normal abundances in most elements except for Li and Na. 
These abundance similarities suggest a common origin for the unevolved Li-rich stars and low-mass metal-poor stars with massive compact companions. 
We also made the first detection of N abundance to unevolved Li-rich stars in \objname, and found that it is significantly enhanced ($\xfe{N}=1.3$).
The observed abundance pattern of \objname, spanning from C to Si, indicates that its surface has been polluted by an intermediate mass former companion star or a nova system that involves a massive ONe white dwarf.
Using a population synthesis model, we show that the nova scenario can lead to the observed level of Li enhancement and also provide an explanation for Li-rich stars without companions and those with massive compact companions.
}

\date{Received ; accepted }

\maketitle
\section{Introduction}

Lithium provides a unique window into the evolution and structure of low-mass stars. 
It is quickly destroyed at temperatures above $\sim 2.5 \times 10^6\,\mathrm{K}$, which can easily be reached in stellar interiors.
This results in a significant decrease in surface Li abundance of low-mass stars at the first dredge-up, when they develop a deep convective envelope that mixes the Li-depleted material from the interior to the surface.
However, it has been known for more than four decades that a small fraction of low-mass red giant branch (RGB) stars show enormous Li abundance and are called Li-rich giants \citep{Wallerstein1982a}.
Thanks to large spectroscopic surveys and the precise identification of their evolutionary status from asteroseismology, our understanding of the origin of such Li-rich giants has improved significantly in recent years \citep{SilvaAguirre2014a,Jofre2015a,Casey2016a,Smiljanic2018a,Casey2019a,Singh2019a,Kumar2020a,Yan2021a,Singh2021a,Deepak2021a,Martell2021a,Zhou2022a,Mallick2023a,Tayar2023a,Castro-Tapia2024a}. 
Theoretical efforts have also been made to explain the origin of enhanced lithium in evolved Li-rich stars \citep{Cameron1971a,Zhang2020a,Schwab2020a,Mori2021a,Gao2022a,Denissenkov2023a,Li2023a}.
Thanks to these studies, we now know that the evolutionary phase near the red clump plays a role in producing most Li-rich stars, although it alone cannot explain all the observed Li-rich giants.

In more recent years, some Li-rich stars have been found also among unevolved stars, which have not yet undergone the first dredge-up.
Such unevolved Li-rich stars are extremely rare (only $<10$ stars confirmed to be Li-rich from high-resolution spectroscopy), but found at low metallicity ([Fe/H] $< -1.0$) in the field \citep[][hereafter L18]{Li2018a} and globular clusters \citep{Koch2011a,Monaco2012a}, and at solar metallicity \citep{Deliyannis2002a,Yan2022a}.
We focus on metal-poor unevolved Li-rich stars in this study as their Li enhancements can clearly be identified thanks the almost constant Li abundance of typical metal-poor stars \citep{Spite1982a,Spite1982b,Charbonnel2005a}\footnote{Accretion of circumstellar object and diffusion have been suggested to be responsible for the Li-richness for solar metallicity, unevolved Li-rich stars.}.
Their Li abundances exceed the prediction from the standard Big Bang nucleosynthesis \citep[e.g.,][]{Coc2002a,Cyburt2016a}, requiring the excess Li to be produced somewhere else.
As they are unevolved, the Li-excess cannot be explained by intrinsic nucleosynthesis and post first dredge-up mixing, making them the most challenging class of Li-rich stars to explain.
Hence, pollution from an external source is discussed as a possible origin of the Li-rich stars in the literature \citep[e.g.,][]{Koch2011a,Li2018a}.
The external source could be planets, highly evolved RGB or asymptotic giant branch (AGB) companions, or novae \citepalias{Li2018a}.
Better characterizing these unevolved Li-rich stars and enlarging the existing sample are of great interest to constrain their origin.
This would then provide insights into the synthesis of Li in the Universe, which is still not fully understood.

Here, we report the discovery of an unevolved Li-rich metal-poor star, Gaia DR3 6334970766103389824 (hereafter \objname).
This star was identified as a member of the stellar stream ED-3 from its kinematics \citep{Dodd2023a} and selected as a target for follow-up high-resolution spectroscopy, to study the stream's progenitor \citep[][hereafter D25]{Dodd2024a}. 
The star has an effective temperature (\teff) of $6417\,\mathrm{K}$ and surface gravity ($\log g$) of $4.27$, and our high-resolution spectroscopic observation reveals it to be a metal-poor star with $\feh=-2.1$ and an extremely high Li abundance of $\ax{Li}=3.5$, which approximately 1.3 dex higher than typical unevolved metal-poor stars \citep[e.g.,][]{Charbonnel2005a}.
We report the detailed abundance pattern of the star, including the first measurement of N and Al abundances for an unevolved Li-rich star, to discuss its possible origins.

We complement this discovery with detailed abundance analysis of three metal-poor main-sequence turn-off stars from \citet[][hereafter E24]{El-Badry2024a}.
These stars are identified as having massive, non-luminous companions, which are candidates for neutron stars.
Interestingly, all three metal-poor stars with $\feh<-1$ in \citetalias{El-Badry2024a} were reported to be Li-rich, which might indicate a common origin for the unevolved Li-rich stars and the metal-poor stars with massive compact companions.
While one of the stars, Gaia DR3 6328149636482597888 (hereafter \gaians), has been extensively studied in \citet{El-Badry2024b}, the other two stars, Gaia DR3 1350295047363872512 and 5136025521527939072 (hereafter \nscanone\ and \nscanfive) have not been subjected to detailed abundance analysis before now.
We investigate the link between these three stars with massive compact companions and the previously identified unevolved Li-rich stars by comparing their abundance patterns.

This paper is organized as follows. 
In Sections~\ref{sec:data} and \ref{sec:analysis}, we describe the data and analysis of the spectra, respectively.
After presenting the results in Section~\ref{sec:results}, we discuss the possible origin of unevolved Li-rich stars in Section~\ref{sec:discussion}.
We then conclude in Section~\ref{sec:conclusion}.

\section{Data \label{sec:data}}

\objname\ was identified as a member of the stellar stream ED-3 from its kinematics and followed up with the Ultraviolet and Visual Echelle Spectrograph \citep[UVES;][]{Dekker2000a} of the Very Large Telescope (program ID: 0111.D-0263(A), P.I. E. Dodd).
The spectrum was obtained with one of the standard setups of UVES, which covers the wavelength range from 328 nm to 683 nm with a resolving power of about 55000.
We used the phase 3 product for further analysis.
We have also used another archival spectrum of the same star obtained with the same instrument one year earlier by \citet{Ceccarelli2024a} under the program ID 0109.B-0522(B) to check for radial velocity variation. 

We analyze the high-resolution spectra used by \citetalias{El-Badry2024a} for the three stars;
the spectrum of \nscanone\ was obtained with the High Resolution Echelle Spectrometer \citep[HIRES;][]{Vogt1994a} on the Keck I telescope, that of \gaians\ was obtained with the Magellan Inamori Kyocera Echelle \citep[MIKE;][]{Bernstein2003a} on the Magellan Clay telescope, and that of \nscanfive\ was obtained with the Tillinghast Reflector Echelle Spectrograph \citep[TRES;][]{Furesz2008a} on the 1.5m Tillinghast telescope. 
The spectral resolutions of the instruments are $R\sim 55,000$, 40,000--55,000, and 44,000, and the wavelength coverages are 365--800 nm, 385--910 nm, and 333--967 nm, respectively.
Readers are referred to \citetalias{El-Badry2024a} for the details of the observations and data reduction.
Note that while these spectra cover a wider wavelength coverage in the red region than the UVES spectrum, we only analyze the region that is in common with the UVES spectrum (328--683 nm) for consistency.

\section{Analysis\label{sec:analysis}}

The analysis mostly follows the method described in \citetalias{Dodd2024a}. 
Below we summarize the analysis and describe differences.
We performed the abundance analysis primarily with the 1D LTE analysis code MOOG \citep{Sneden1973a} unless otherwise specified. 
For \objname, the stellar parameters were estimated from the Gaia and 2MASS photometry using the color-\teff\ relation of \citet{Mucciarelli2021a}, the bolometric correction of \citet{Casagrande2014a}, and the extinction map of \citet{Lallement2022a}.
The microturbulent velocity ($\xi$) is estimated by minimizing the trend between Fe abundances derived from individual \ion{Fe}{I} lines and their reduced equivalent widths.
We also validate the \teff\, with a spectroscopic method based on a differential abundance analysis using the python package \texttt{q2} \citep{Ramirez2014a}.
It utilizes the excitation equilibrium of \ion{Fe}{I} lines, using HD84937 as the reference star, for which we adopt $\teff=6356\,\mathrm{K}$ \citep{Heiter2015a}, $\logg = 4.13$ \citep{Giribaldi2021a}, $\xi=1.39\,\kms$ \citep{Jofre2015b}, and $\xh{Fe}=-1.97$ \citep{Amarsi2022a}.
We also confirmed that the H$\alpha$ wings are consistent with $\teff\sim 6400\,\mathrm{K}$ using the grid of synthetic spectra from \citet{Amarsi2018a}\footnote{We note that we re-reduced the raw UVES data to check the blaze function using the UVES data reduction pipeline version 6.4.6 \citep{Ballester2000a}.}.
The \teff\ and $\log g$ for the \citetalias{El-Badry2024a} stars were taken from \citetalias{El-Badry2024a} but the microturbulent velocity and metallicity were rederived by the same method as for \objname.

Abundances are derived from equivalent widths for Fe and spectral synthesis for Li, C, N, Na, Mg, Al, Si, Mn, Cu, Sr, Y, and Ba, La, and Eu, with hyperfine structure splitting considered for Na, Al, Mn, Ba, and Eu. 
For the Li lines, we analyzed both 6104 and 6708 $\mathrm{\AA}$ lines and used interpolated 3D non-LTE synthetic spectra from \textsc{Breidablik} \citep{Wang2021a,Wang2024a} instead of MOOG. 
Abundances of Na and Al are corrected for non-LTE effects using the grid of \citet{Lind2022a}.
Table~\ref{tab:param_abun} summarizes measured stellar parameters and abundance of key elements as well as the properties of the star, and the abundances of the three stars from \citetalias{El-Badry2024a} are summarized in Table~\ref{tab:param_El-Badry}.
The complete information on the linelist and abundances other than those reported in Table~\ref{tab:param_abun} are provided in \citetalias{Dodd2024a}.

The abundances of \gaians\ has been reported in \citet{El-Badry2024b}.
We confirm that our abundance measurements are largely consistent with theirs; for 17 elements in common, the mean difference is $0.05$ dex in \xfe {X} with a standard deviation of 0.15 dex.
A notable difference is found for a few elements, namely Na, Co, and Eu. 
Our Na abundance is significantly lower ($\xfe{Na}=0.00$) than theirs ($\xfe{Na}=0.24$), which could partly be due to the non-LTE correction of 0.11 dex applied in this study. 
The Co abundance in this study is based on just one line thus the abundance is less reliable than the other elements.
The largest difference is seen in Eu abundance; our Eu abundance is higher ($\xfe{Eu}=0.49$) than \citet{El-Badry2024b} ($\xfe{Eu}=0.09$).
While the source of this difference is unclear, the abundance pattern of \gaians\ is now more consistent with the $r$-process dominated pattern of neutron-capture elements. 
This is more common in metal-poor stars without significant enhancements in $s$-process elements, such as Ba and La. 

The radial velocity of \objname\ is measured during the abundance analysis by comparing the observed wavelengths of Fe lines with the laboratory wavelengths.
Although we obtain an uncertainty of 0.05 \kms\ from the scatter in this comparison, this does not include the systematic uncertainty of the wavelength calibration and would likely be an underestimate (see the discussion in the next Section).

\begin{table}
    \caption{The property of \objname\label{tab:param_abun}}
    \begin{tabular}{ll}
        \hline \hline
        Property & Value \\
        \hline
        Gaia DR3 ID & 6334970766103389824 \\
        $v_r$ (Gaia DR3) & $(166.98\pm 3.03)\, \kms$ \\
        $v_r$ (VLT, 2023-08-26) & 170.55 \kms \\
        $v_r$ (VLT, 2022-05-30) & 170.91 \kms \\ 
        \hline
        \teff ($G-K_s$) & $(6417\pm 90)\,\mathrm{K}$ \\
        \logg (Photo) & $4.27\pm 0.05$ \\
        \vmic & $(1.44\pm 0.12)\,\kms$ \\
        \teff (Spec) & $(6333\pm82)\,\mathrm{K}$\\
        \logg (Spec) & $4.03\pm 0.10$\\
        \hline 
        $\xh{Fe}_{\rm I}$ & $-2.22\pm 0.07$\\ 
        $\xh{Fe}_{\rm II}$ & $-2.06\pm 0.03$\\ 
        \ax{Li} & $3.46\pm 0.11$ \\
        \xfe{C} & $0.39\pm 0.13$\\
        \xfe{N} & $1.30\pm 0.13$\\
        \xfe{Na} & $1.04\pm 0.10$\\
        \xfe{Mg} & $0.28\pm 0.03$\\
        \xfe{Al} & $-0.32\pm 0.07$\\
        \xfe{Si} & $0.10\pm 0.10$\\
        \xfe{Sr} & $-0.18\pm 0.11$\\
        \xfe{Y}  & $-0.03\pm 0.10$\\
        \xfe{Ba} & $-0.37\pm 0.07$\\
        \hline
    \end{tabular}
    \end{table}

\begin{table*}
\caption{Stellar parameters and abundances of \citetalias{El-Badry2024a} stars. \label{tab:param_El-Badry}}
\begin{tabular}{l*{9}{c}}
\hline \hline
Param                    & \multicolumn{3}{c}{E24\_1350} & \multicolumn{3}{c}{E24\_5136} & \multicolumn{3}{c}{E24\_6328} \\
& value & error & $N_{\rm line}$ & value & error & $N_{\rm line}$ & value & error & $N_{\rm line}$ \\
\hline
\teff \tablefootmark{a} (K)   & 6289 & 32   & & 6470 & 40  && 6049 & 23  &\\
\logg \tablefootmark{a}       & 4.02 & 0.01 & & 4.19 & 0.01&& 3.99 & 0.01&\\
\vmic ($\mathrm{km\,s^{-1}}$) & 1.33 & 0.11 & & 1.33 & 0.17&& 1.60 & 0.12&\\
A(Li) \tablefootmark{a} & 3.53 & 0.09 & 1 & 3.11 & 0.08 & 1 & 2.90 & 0.08 & 1 \\
$\xh{Fe}_{\rm I} $  &  -1.57 &   0.02 & 107 &  -1.35 &   0.03 & 99 &  -1.47 &   0.10 & 140 \\
$\xh{Fe}_{\rm II} $  &  -1.53 &   0.03 & 10 &  -1.37 &   0.03 & 9 &  -1.45 &   0.11 & 11 \\
$\xfe{C} $  &   0.84 &   0.12 & 1 &   0.48 &   0.16 & 1 &  -0.40 &   0.15 & 1 \\
$\xfe{N} $  & & & & & & &   0.55 &   0.14 & 1 \\
$\xfe{Na} $  &   0.84 &   0.11 & 1 &   0.62 &   0.15 & 1 &  -0.00 &   0.15 & 1 \\
$\xfe{Mg} $  &   0.30 &   0.06 & 5 &   0.23 &   0.05 & 5 &   0.23 &   0.10 & 5 \\
$\xfe{Al} $  &  -0.35 &   0.11 & 1 &  -0.69 &   0.16 & 1 &  -0.57 &   0.10 & 2 \\
$\xfe{Si} $  &   0.27 &   0.07 & 3 &   0.27 &   0.11 & 2 &   0.30 &   0.10 & 4 \\
$\xfe{Ca} $  &   0.29 &   0.03 & 17 &   0.29 &   0.02 & 18 &   0.36 &   0.03 & 21 \\
$\xfe{Sc} $  &   0.11 &   0.03 & 7 &   0.12 &   0.11 & 5 &   0.17 &   0.05 & 7 \\
$\xfe{Ti}_{\rm I} $  &   0.38 &   0.03 & 9 &   0.45 &   0.06 & 6 &   0.27 &   0.07 & 8 \\
$\xfe{Ti}_{\rm II} $  &   0.40 &   0.03 & 22 &   0.64 &   0.16 & 21 &   0.40 &   0.06 & 22 \\
$\xfe{V} $  &  -0.08 &   0.11 & 1 & & & &   0.04 &   0.10 & 3 \\
$\xfe{Cr} $  &  -0.06 &   0.03 & 9 &   0.04 &   0.06 & 6 &  -0.09 &   0.03 & 9 \\
$\xfe{Mn} $  &  -0.17 &   0.11 & 4 &  -0.02 &   0.10 & 3 &  -0.36 &   0.09 & 8 \\
$\xfe{Co} $  &   0.14 &   0.07 & 2 & & & &   0.19 &   0.14 & 1 \\
$\xfe{Ni} $  &  -0.05 &   0.06 & 6 &  -0.07 &   0.03 & 5 &  -0.00 &   0.07 & 10 \\
$\xfe{Zn} $  &   0.29 &   0.10 & 1 &   0.27 &   0.15 & 1 &   0.09 &   0.10 & 2 \\
$\xfe{Sr} $  &   0.19 &   0.09 & 2 &   0.29 &   0.13 & 2 &  -0.08 &   0.10 & 2 \\
$\xfe{Y} $  &  -0.12 &   0.08 & 2 &  -0.03 &   0.15 & 1 &   0.00 &   0.10 & 4 \\
$\xfe{Zr} $  &   0.31 &   0.08 & 2 & & & &   0.19 &   0.13 & 2 \\
$\xfe{Ba} $  &   0.01 &   0.08 & 2 &   0.11 &   0.10 & 4 &  -0.07 &   0.12 & 4 \\
$\xfe{La} $  & & & & & & &   0.28 &   0.15 & 1 \\
$\xfe{Eu} $  &   0.59 &   0.12 & 1 & & & &   0.49 &   0.14 & 2 \\
\hline
\end{tabular}
\tablefoot{
\tablefoottext{a}{Taken from \citetalias{El-Badry2024a}.}
}
\end{table*}

\section{Results\label{sec:results}}
\begin{figure}
    \centering
    \includegraphics[width=1.0\hsize]{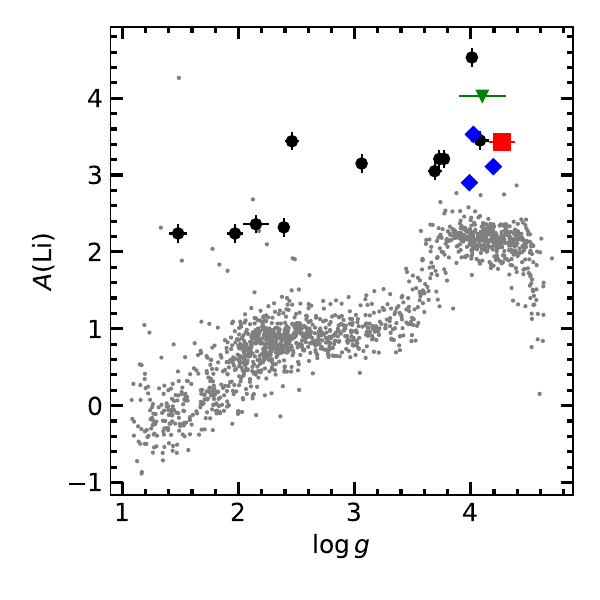}
    \caption{Li abundance as a function of \logg. The red square shows \objname, the green triangle shows the Li-rich star in NGC 6397 \citep{Koch2011a}, the blue diamonds show the three metal-poor stars in \citetalias{El-Badry2024a}, and the black circles show the Li-rich stars from \citetalias{Li2018a}. The grey dots are stars with $\feh<-1.3$ from the GALAH survey \citep{Buder2021a,Wang2024a}. \label{fig:Li_logg}}
\end{figure}

Both spectroscopic and photometric methods suggest that \objname\ is around the main-sequence turn-off point and has not experienced the first dredge-up. 
\fig{fig:spectra} shows a portion of the spectrum of \objname. 
Both Li lines are well-fitted by a Li abundance of $\ax{Li}=3.46$, which is about 1.3 dex higher than in typical metal-poor stars (\fig{fig:Li_logg}), which makes this star join the group of unevolved Li-rich stars.
It has the lowest surface gravity among the previously known unevolved Li-rich stars and hence can be considered to be the least evolved star.
Most elements, including those of C, Al, $s$-process elements, such as Sr and Ba, show typical abundance of metal-poor stars at similar metallicity. 
The exceptions are N and Na; their abundances are $\xfe{N}=1.3$ and $\xfe{Na}=1.04$, respectively.
As far as we know, we determined N abundance for the first time in an unevolved Li-rich star. 
\fig{fig:spectra} confirms that the strength of NH lines is consistent with the high N abundance ratio.

\nscanone\ and \nscanfive, show clear Na enhancement.
The NH feature, which we used for N abundance measurement for \objname, is not covered in the spectra of these two stars.
\gaians\ shows a slight C depletion, slight N enhancement, and subtle Na enhancement compared to typical metal-poor stars, but each of these signatures is subtle, and does not represent a significant deviation from typical metal-poor stars.
We do not detect any abundance anomalies in other elements for the three stars.

\begin{figure*}
    \includegraphics[width=\hsize]{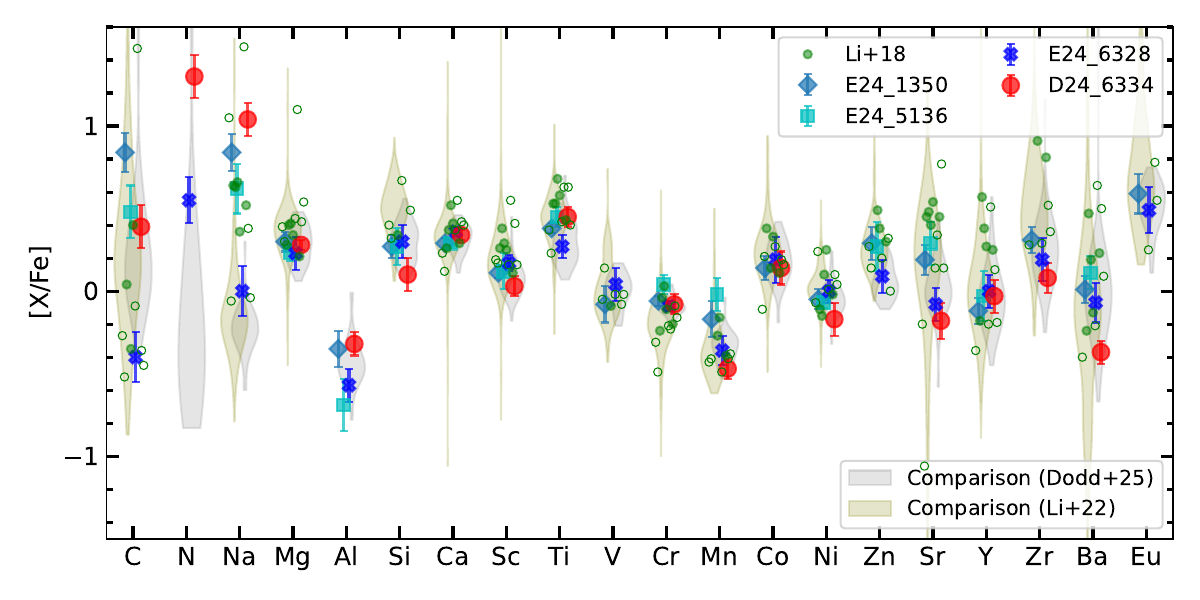}
    \caption{Abundance patterns of \objname, the three stars from \citetalias{El-Badry2024a}, and Li-rich stars from \citetalias{Li2018a}, compared to typical metal-poor stars, which are shown as violin plots. One of the comparison samples is taken from \citetalias{Dodd2024a}, in which abundances are derived consistently with the present study. The other comparison sample is taken from \citet{Li2022a}, which mostly contains stars with $\feh<-2$. Note that Li-rich stars in \citetalias{Li2018a} are reanalyzed by \citet{Li2022a}, and we use the latter abundance for consistency. Five unevolved Li-rich stars in \citet{Li2018a} are shown with filled symbols, and the other more evolved Li-rich stars are shown with open symbols. \label{fig:abun_pattern}
    }
\end{figure*}

These abundance features of \objname\ and three \citetalias{El-Badry2024a} stars are consistent with the properties of the six previously identified, unevolved Li-rich stars as shown in \fig{fig:abun_pattern} \citep{Koch2011a,Pasquini2014a,Li2018a,Li2022a}.
\objname, \nscanone, and \nscanfive\ show clear Na enhancement similar to most of Li-rich stars, while the Na enhancement in \gaians\ is tenuous, which is also seen in a minor fraction of Li-rich stars. 
There has been no measurement of N abundance for unevolved Li-rich stars in the literature, and hence we can not conclude if the N enhancement seen in \objname\ is common among unevolved Li-rich stars.  

While the three stars from \citetalias{El-Badry2024a} are known to have massive non-luminous companions, other known unevolved Li-rich stars do not necessarily show radial velocity variation \citep{Koch2012a,Li2018a}.
There is no evidence for the presence of a companion star in the radial velocity measurements of \objname.
The relative radial velocity between the two UVES observations is estimated to be 0.36 \kms\ by cross-correlating the two spectra.
This is likely to be insignificant compared to the systematic uncertainty, which can be up to $\sim 0.5 \kms$ \citep{Whitmore2010a}.
Further, we note that the telluric lines in the two spectra are shifted by around 0.5 \kms. 
While all the currently available measurements seem to be consistent with each other, future multi-epoch spectroscopy and astrometry that will be available following Gaia DR4 will provide a more robust conclusion on the binarity of \objname.
For now, we simply note \objname\ is similar to the majority of the known unevolved Li-rich stars in the sense that they do not show radial velocity variation so far \citep{Li2018a,Aoki2022a}.

We also note that there is no indication of ongoing stellar activity in the spectrum of \objname.
None of emission lines, fast rotation, nor infrared excess are detected for the object.

\section{Discussion\label{sec:discussion}}

\begin{figure*}
    \centering
    \includegraphics[width=1.\hsize]{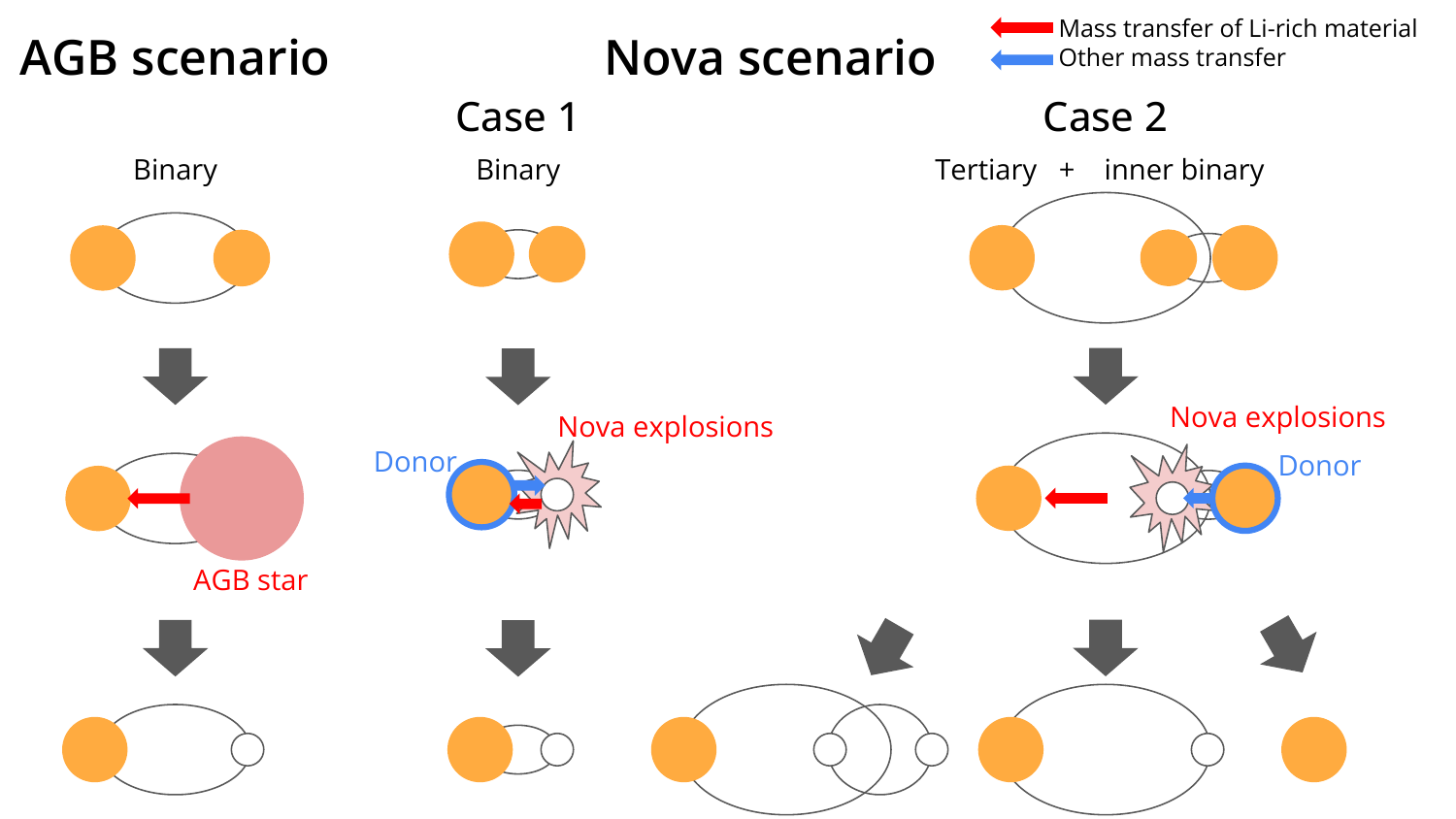}
    \caption{Schematic diagrams of formation scenarios of unevolved Li-rich stars considered in this study. In the AGB scenario shown in the leftmost column, the pollution occurs from a giant star to the currently observed star. The donor star is now expected to be a white dwarf. The remaining columns show scenarios in which pollution from a classical nova is considered. Depending on whether the currently observed star was the donor or not, we further divide the nova scenario into cases 1 and 2. The final state of the nova scenario can be a single, binary, or triple-star system. \label{fig:schematic}}
\end{figure*}

\begin{figure*}
    \centering
    \includegraphics[width=1.\hsize]{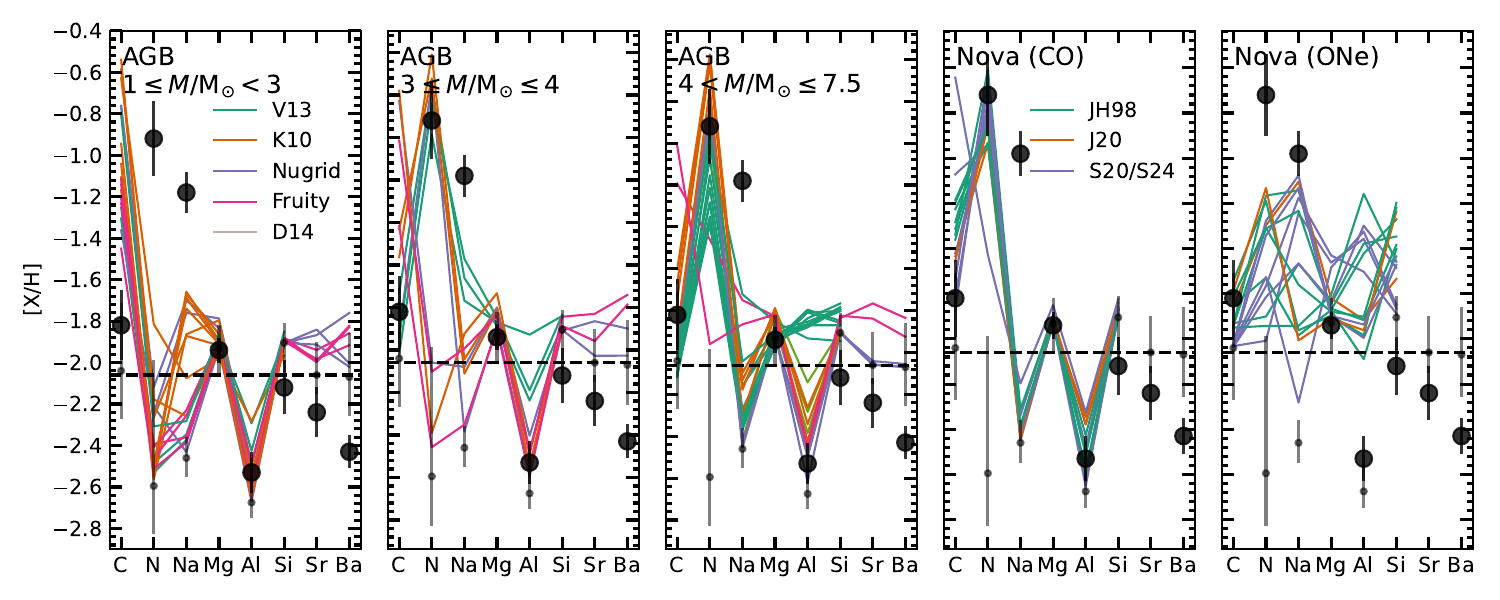}
\caption{Abundance pattern of \objname\ (black points) compared to the theoretical chemical yields from AGB stars from \citet[][V13]{Ventura2013a}, \citet[][FRUITY]{Cristallo2015a}, \citet[][K10]{Karakas2010a}, \citet[][NuGrid]{Ritter2018a}, and \citet[][D14]{Doherty2014a} and those from nova models from \citet[][JH98]{Jose1998a}, \citet[][J20]{Jose2020a}, \citet[][S20]{Starrfield2020a}, and \citet[][S24]{Starrfield2024a}. The small gray dots show abundances that a star with the same metallicity as \objname\ would have if it had \xfe{X} values equal to the median values of all the stars in \citetalias{Dodd2024a}. Yield from each model is dilluted with the non-polluted abundance (see text). 
\label{fig:pattern}}
\end{figure*}

\begin{figure*}
    \centering
    \begin{minipage}{1.0\textwidth}
        \centering
        \includegraphics[width=1.\hsize]{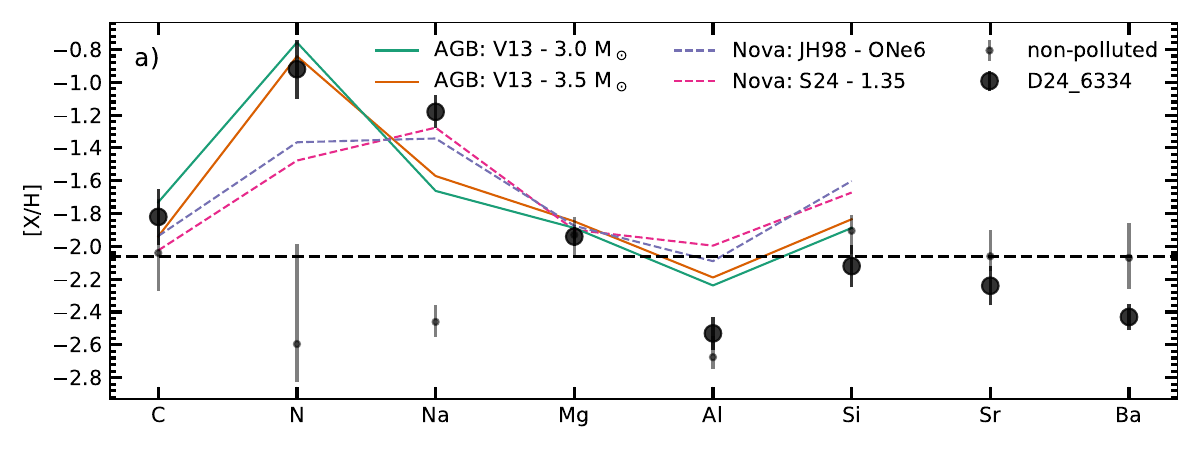}
    \end{minipage}\\

    \begin{minipage}{1.\textwidth}
    \centering
    \includegraphics[width=1.\hsize]{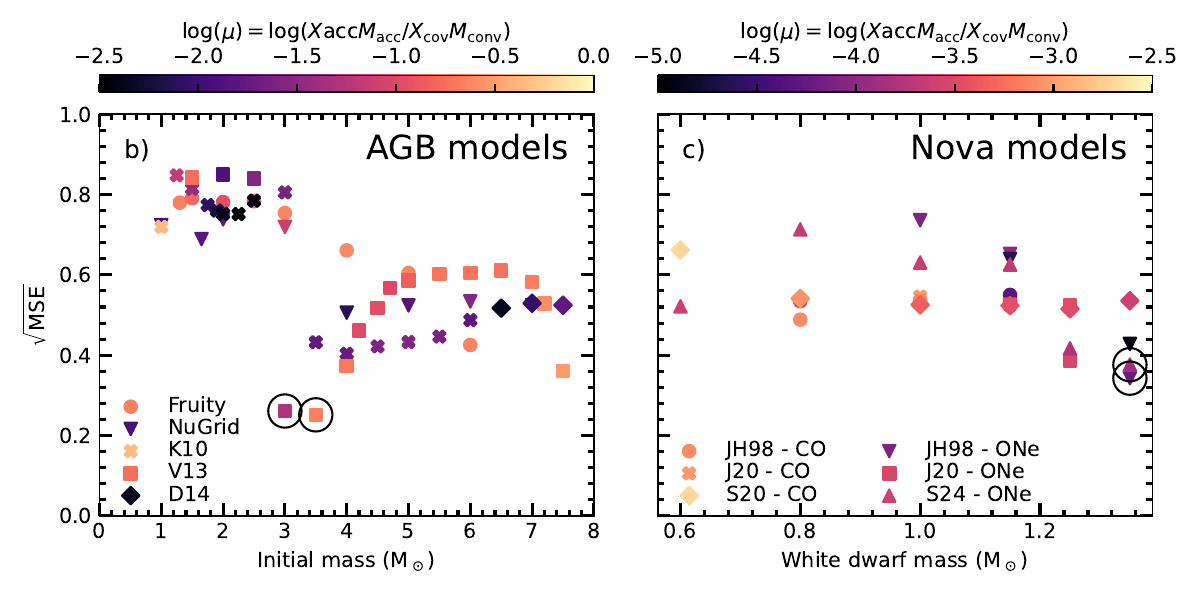}
    \end{minipage}
        
    \caption{Summary of the comparison between the observed abundance pattern of \objname\ and the theoretical chemical yields from AGB stars and nova models. Panel a) is the same as \fig{fig:pattern} but only for four selected models. Panels b) and c) summarize the square root of the sum of the mean squared error for each model for AGB scenario (b) and for nova scenario (c).  The color indicates the dilution factor, defined as the ratio of the hydrogen mass of accretion origin to the total hydrogen mass in the envelope after the accretion (see text). The four models selected in panel a) are highlighted with open circles. They are $3.0\,\mathrm{M_{\odot}}$ and $3.5\,\mathrm{M_{\odot}}$ AGB models from \citet{Ventura2013a}, and ``ONe6'' nova model from \citet{Jose1998a} and $1.35\,\mathrm{M_{\odot}}$ nova model from \citet{Starrfield2024a}.
    \label{fig:pattern_summary}
    }
    
\end{figure*}

The origin of Li-rich stars is still an open question. 
One of the proposed scenarios is the production of Li in the observed star itself during a certain evolutionary phase near the red clump phase. 
While this could be the case for the Li-rich stars in the red clump phase \citep{Schwab2020a,Mori2021a,Gao2022a,Denissenkov2023a,Li2023a}, it is unlikely for unevolved Li-rich stars, which have not undergone even the first dredge-up.
Other scenarios include engulfment of planets or planetesimal material \citep[e.g.,][]{Alexander1967a,Siess1999a}, transfer of Li-rich material from an evolved companion, Li production inside the star and transport of the produced Li to the surface through merger-induced mixing \citep{Kravtsov2024a}. 

In this section, we interpret the abundance pattern of \objname\ and attempt to narrow down the origin of unevolved Li-rich stars, assuming that the enhanced Li and other features of \objname\ in its abundance pattern are related and that its abundance patterns are representative of unevolved Li-rich stars.
Since the star is unevolved, it is relatively safe to assume that the surface chemical composition has not been altered by the evolution of the star itself since its formation or the event that led to the Li enhancements, allowing us to use all the elements, including C and N to constrain the formation mechanism. 
We have also shown that at least two of the three metal-poor stars from \citetalias{El-Badry2024a} show abundance patterns similar to \objname\ and other known unevolved Li-rich stars, suggesting that the formation models need to be able to explain the presence of Li-rich stars without companions and those with massive compact companions.

As discussed in \citetalias{Li2018a}, the planet engulfment scenario is unlikely for metal-poor Li-rich stars since planets are expected to be rare in metal-poor environments \citep[e.g.,][]{Wang2015a,Andama2024a}.
The high \xfe{N} and \xfe{Na} abundances are also difficult to explain by this scenario since they have lower condensation temperature than Li \citep{Lodders2003a}.
The merger-induced nucleosynthesis and mixing seems also unlikely since the production of C, which requires a high temperature ($\gtrsim 10^8\,\mathrm{K}$), would be needed to explain the enhanced C+N abundance of $\xfe{C+N}=0.72$.

The remaining scenario is the transfer of Li-rich material from (a) companion(s). 
We here discuss two possibilities: pollution from a giant star and pollution from a classical nova (\fig{fig:schematic}).
Below the abundance pattern of \objname\ is compared to the theoretical chemical yields of giant star models and nova models to see if either of these scenarios can explain the observed abundance. 
We express the predicted abundance pattern as a mixture of material with a typical abundance pattern of metal-poor stars and the theoretical chemical yields of nucleosynthesis in a giant and nova.
We define the non-polluted abundance in [X/H] ($\xh{X}_0$) using the median \xfe{X} values of all 41 stars in \citetalias{Dodd2024a} and the metallicity of \objname.
The theoretical chemical yields in [X/H] ($\xh{X}_{\rm model}$) are obtained from [X/Fe] \citep{Ventura2013a,Cristallo2015a}, overproduction factors \citep{Karakas2010a,Doherty2014a}, or yields \citep{Ritter2018a} assuming $\feh=-2.1$ for giant star models and from the mass fraction of elements in ejecta for nova models \citep{Jose1998a,Jose2020a,Starrfield2020a,Starrfield2024a}. 
We then fit the observed [X/H] of \objname\ ($\xh{X}_{\rm target}$) by minimizing the following mean squared error:
{\small
\[
MSE(\mu) = \sum_{i} \left(
\xh{X}_{i,\rm target}-\log(\mu 10^{\xh{X}_{i, \rm model}}+(1-\mu)10^{\xh{X}_{i,0}})
\right) ^2,
\]}
where $\mu$ is the dilution factor with $0<\mu<1$, which is the ratio of the hydrogen mass of accretion origin ($X_{\rm acc}M_{\rm acc}$) to the total hydrogen mass after the accretion in the convective envelope of \objname\ ($X_{\rm env}M_{\rm env}$).
We use abundances of C, N, Na, Mg, Al, and Si as these are provided in all the models considered here.
Li is not included in this fitting because theoretical prediction tends to be less uncertain, but we discuss if the observed Li abundance of \objname\ and those of typical unevolved Li-rich stars can be reproduced by the combination of $\mu$ and the expected range of Li production.
We also note that $s$-process elements are not included either since they are not provided in every model, but we do compare the predicted Sr and Ba abundances with the observation when available.

\subsection{Pollution from a giant star}
A comparison to asymptotic giant branch (AGB) star models is of particular interest since Li production is suggested in AGB stars \citep{Sackmann1992a}.
The high C+N abundance also requires the companion to be an AGB star, as C needs to be synthesized through He-burning and some of the produced C needs to be converted to N before the mass transfer event.   
We thus consider AGB models from \cite{Karakas2010a}, \citet{Ventura2013a}, \citet[][FRUITY]{Cristallo2015a}, and \citet[][NuGrid]{Ritter2018a} and super AGB models from \citet{Doherty2014a} and compare them to the observed abundance pattern of \objname.
Depending on the availability, we chose models that match the metallicity of \objname, $\feh\sim -2.1$:$Z=0.0003$ for \citet{Ventura2013a} and \citet{Cristallo2015a} and $Z=0.0001$ for \citet{Karakas2010a}, \citet{Ritter2018a} and \citet{Doherty2014a}. 
The results of the fitting are shown in \fig{fig:pattern} and summarized in \fig{fig:pattern_summary}.

Although no single model of AGB stars can fit the observed abundance pattern of \objname\ perfectly, we can still obtain some insights from the comparison.
Except for the FRUITY model, high \xfe{N} and [N/C] ratios are realized in intermediate-mass AGB stars with the initial mass of $M_{\rm ini}\gtrsim 3\,\mathrm{M_\odot}$ as a result of hot bottom burning. 
Such models also predict high Na abundances at $M_{\rm ini}\lesssim 5\,\mathrm{M_\odot}$.
This may suggest that the most promising companion is an AGB star with $M_{\rm ini} \sim 4\,\mathrm{M_\odot}$.
Among such intermediate mass AGB models, $3\,\mathrm{M_\odot}$ and $3.5\,\mathrm{M_\odot}$ models from \citet{Ventura2013a} particularly show good fits to the observed abundance pattern of \objname\ (\fig{fig:pattern_summary}).
While such AGB stars are expected to produce some $s$-process elements, such as Sr and Ba, $s$-process enhancements might be at most subtle after the dilution (see FRUITY and Nugrid model predictions in \fig{fig:pattern}).

In the two best-fitting models from \citet{Ventura2013a}, the accreted mass consists of about 10\% of the final convective envelope. 
Theoretical predictions show that the surface Li abundance of AGB stars can be $\ax{Li}\sim 4-5$ \citep[e.g.,][]{Choplin2024a}, which is also confirmed by some observations \citep[e.g.,][]{Abia1997a,Abia2000a}.
The observed Li abundance of most unevolved Li-rich stars ($\ax{Li}\sim 3$) can easily be reproduced with the obtained dilution factor and the expected Li yield from AGB stars. 

While the two models from \citet{Ventura2013a} might seem promising from the arguments above, the behaviour of AGB yields with initial mass is not straightforward and highly model-dependent as seen in Figures~\ref{fig:pattern} and \ref{fig:pattern_summary}.
Thus we cannot robustly conclude that pollution from an intermediate mass AGB star is the origin of the Li-rich stars.
Further refinements and observational constraints on the AGB evolution is highly desired to reduce the uncertainty in the AGB yields and to make a more robust conclusion.


In the scenario of pollution from an intermediate mass AGB star, the presence of a white dwarf companion with $\sim 0.8-0.9\,\mathrm{M_\odot}$ is expected \citep{Ventura2013a}.
\objname, however, shows no observational evidence for the presence of a white-dwarf companion.
The RUWE in Gaia DR3, which is an indicator of deviation in astrometry from a single star solution and can be diagnostic for the presence of companions \citep{Penoyre2020a,Belokurov2020a,Castro-Ginard2024a}, is small, and there is no variation in radial velocity.
If it does not currently have a companion, it might have lost the companion through a dynamical interaction with other objects, or it might have been born from a gas cloud that was rich in AGB ejecta.
Future epoch astrometry from the \textit{Gaia} mission and long-term monitoring of radial velocity variation would be needed to put a strong constraint on the presence of a companion. 
\citetalias{Li2018a} also reported that the majority of unevolved Li-rich stars do not show radial velocity variation, which implies that \objname\ is not an exception.
Moreover, Li-rich stars known to have companions from \citetalias{El-Badry2024a} have companion masses of $>1.2\,\mathrm{M_\odot}$, which is inconsistent with the companion mass expected for an intermediate mass AGB star ($0.8-0.9\,\mathrm{M_{\odot}}$).
Massive super AGB stars that would leave a massive white dwarf companion do not produce enough Na and are unable to explain the observed Na enhancements (\fig{fig:pattern}). 
These inconsistencies between the observations and the expectations is a challenge to the AGB pollution scenario, motivating us to consider other scenarios.

\subsection{Pollution from a classical nova}

Novae are known to produce Li from both theoretical studies \citep{Arnould1975a,Starrfield1978a,Boffin1993a,Hernanz1996a,Jose1998a,Denissenkov2014a,Starrfield2020a,Jose2020a,Denissenkov2021a,Starrfield2024a} and observational studies \citep{Tajitsu2015a,Izzo2015a,Molaro2016a,Tajitsu2016a,Izzo2018a,Selvelli2018a,Molaro2020a,Arai2021a,Molaro2022a}.
We note here that models tend to underestimate Li production by an order of magnitude compared to the observations \citep[see discussions in][]{Kemp2022a,Gao2024a} and we mainly adopt the observationally constrained Li yield in the following discussion.
Pollution from a nova has been proposed in the literature as an origin of excess-Li in Li-rich giants, but it does not seem to be the dominant formation channel of Li-rich giants \citep{Gratton1989a,Casey2019a}. 
However, the viability of this nova pollution scenario for unevolved Li-rich stars has only briefly been discussed in \citetalias{Li2018a} and not been explored in detail. 
Here, we use abundance ratios of unevolved Li-rich stars, theoretical and observational constraints on the nova nucleosynthesis, and a population synthesis model of novae to study the viability of the nova channel on unevolved Li-rich stars.

In Figures~\ref{fig:pattern} and \ref{fig:pattern_summary}, we show the fitting results of the nova models of \citet{Jose1998a}, \citet{Jose2020a}, \citet{Starrfield2020a}, and \citet{Starrfield2024a} to the observed abundance pattern of \objname \footnote{We used the 25\%--75\% mixing recipe for \citet{Starrfield2020a} and \citet{Starrfield2024a}.}. 
The ONe6 model from \citet{Jose1998a} shows the best match among the nova models considered here with $\mu\sim 10^{-4.2}$. 
The model has a white dwarf mass of $1.35\,\mathrm{M_\odot}$ and the degree of mixing between the core and the envelope is 50\%. 
The second-best model is also a nova by a massive ONe white dwarf; the $1.35\,\mathrm{M_\odot}$ ONe white dwarf model of \citet{Starrfield2024a} fits the abundance pattern of \objname\ with $\mu\sim 10^{-3.9}$.
The match between the observation and these massive ONe white dwarf models is better than most of the AGB models; the exceptions are just the $3\,\mathrm{M_\odot}$ and $3.5\,\mathrm{M_\odot}$ models from \citet{Ventura2013a}, where the fit is similarly good.
Since the observations of novae suggest $\ax{Li}\sim 7$ \citep{Kemp2022a}, the obtained dilution factors would lead to $\ax{Li}\sim 3$, which is in line with the observed Li abundance of \objname.

Thus we consider the nova pollution scenario is as promising as the AGB pollution scenario in terms of purely reproducing the abundance pattern.
The main differences between the intermediate-mass AGB models and massive ONe white dwarf novae are the insufficient Na production in the former and the insufficient N production in the latter. 
Both models predict mild enhancements in Al, which is not observed in \objname.
We note that the nova models are calculated for solar metallicity, and the nucleosynthesis in novae at low metallicity has not been extensively studied.
According to \citet{Jose2007a}, N/C and Na/Al ratios increase at lower metallicity, which is promising for reproducing the abundance pattern.

A nova system needs at least one white dwarf that accumulates material from the companion donor star to explode.
One can consider two cases for a star to be polluted by nova ejecta and to be Li-rich (see \fig{fig:schematic}): 1. the polluted star itself is the donor to the white dwarf, and 2. the star is an outer tertiary to a close binary system that underwent nova explosions. 
Given that the Li-rich stars considered in the present study are unevolved and hence have small radii and negligible mass-loss rates through winds, case 1 implies a close binary system transferring mass through Roche-lobe overflow. 
On the other hand, case 2 does not have any constraints on the separation for the occurrence of novae, but introduces dynamical stability constraints and increases the difficulty in accreting nova ejecta due to increased separation from the accreting white dwarf. 
In both cases, the distance between the nova and the Li-rich star has to be small enough for sufficient pollution to occur. 

The total mass of $^7$Li in the convective envelope of a Li-rich star ($M_{\rm env}(\rm Li)$) can be expressed in terms of its current surface Li abundance ($\ax{Li}$), the mass of the convective envelope ($M_{\rm env}$, typically $10^{-3}\,\mathrm{M_\odot}$ for a metal-poor turn-off star \citep[e.g.,][]{Richard2002a,Matrozis2016a}, and the mass fraction of hydrogen ($X$):
\begin{align}
\displaystyle M_{\rm env}({\rm Li}) &= 10^{\ax{Li}-12}XM_{\rm env}\frac{m(\rm ^7Li)}{m(\rm ^1H)} \nonumber\\
&=5\times 10^{-12}10^{\ax{Li}-3}\left(\frac{X}{0.7}\right)\left(\frac{M_{\rm env}}{10^{-3}\,\mathrm{M_\odot}}\right)\,\mathrm{M_\odot}.\label{eq:needed}
\end{align}
Since the initial Li abundance before the pollution is expected to be around the Spite plateau value ($\ax{Li}\sim 2.2$) \citep[e.g.,][]{Spite1982a,Spite1982b,Charbonnel2005a,Wang2024a}, about 85\% of $M_{\rm env}({\rm Li}$) needs to be accreted by the star to be Li-rich. 
The mass of Li produced in a single nova explosion ($M_{\rm nova}({\rm Li})$) can be expressed with the ejecta mass ($M_{\rm ej}$, typically $10^{-5}\,\mathrm{M_\odot}$; \citealt{Jose1998a}) and the Li mass fraction in it ($X_{\rm Li}$). 
With the observed $^7$Li or $^7$Be abundance of classical nova of $A(^7\mathrm{Li})\sim 7$ \citep[see discussions presented in][]{Kemp2022a,Gao2024a} and the typical mass fraction of hydrogen predicted to be $X\sim 0.3$ from nova simulations \citep{Jose1998a}, the mass fraction of Li in the ejecta is expected to be around $X_{\rm Li}\sim 2\times 10^{-5}$. Thus,
\begin{align}
\displaystyle M_{\rm nova}({\rm Li}) &= X_{\rm Li}M_{\rm ej}\nonumber\\
&=2\times 10^{-10} \left(\frac{X_{\rm Li}}{2\times 10^{-5}}\right)\left(\frac{M_{\rm ej}}{10^{-5}\,\mathrm{M_\odot}}\right)\,\mathrm{M_\odot}.
\end{align}
Considering the high nova ejecta velocities, we assume a purely geometric accretion scenario based on the radius of the polluted star and its distance to the nova. 
The fraction of nova ejecta accreted by the polluted star, $f_{\rm acc}$ is
\begin{equation}
\displaystyle f_{\rm acc} = \frac{\pi R_\star^2}{4\pi d^2},
\end{equation}
where $R_\star$ is the radius of the star, and $d$ is the distance between the nova and the star.
Hence, the total amount of Li that is accreted by the star from $N_{\rm nova}$ nova explosions can be expressed as 
\begin{align}
\displaystyle M_{\rm acc}({\rm Li}) &= N_{\rm nova}M_{\rm nova}({\rm Li})f_{\rm acc} \\
&= 5\times 10^{-11} N_{\rm nova}\left(\frac{X_{\rm Li}}{2\times 10^{-5}}\right)\left(\frac{M_{\rm ej}}{10^{-5}\,\mathrm{M_\odot}}\right)\left(\frac{R_\star}{\mathrm{R_\odot}}\right)^2\left(\frac{d}{\mathrm{R_\odot}}\right)^{-2}\,\mathrm{M_\odot},\label{eq:case1}
\end{align}
or 
\begin{equation}
    \displaystyle M_{\rm acc}({\rm Li})= 1\times 10^{-15} N_{\rm nova}\left(\frac{X_{\rm Li}}{2\times 10^{-5}}\right)\left(\frac{M_{\rm ej}}{10^{-5}\,\mathrm{M_\odot}}\right)\left(\frac{R_\star}{\mathrm{R_\odot}}\right)^2\left(\frac{d}{\mathrm{au}}\right)^{-2}\,\mathrm{M_\odot}. \label{eq:case2}
\end{equation}
Note that these two conditions are the same but are expressed in different units for $d$.
For a star to be Li-rich, 
\begin{equation}
    M_{\rm env}({\rm Li})>M_{\rm acc}({\rm Li})  \label{eq:condition}
\end{equation}
has to be met.

Based on these estimates, we discuss cases 1 and 2 in the following using the nova population synthesis model of \citet{Kemp2021a}.
Since we expect small $d$, $\sim$ a few $R_{\odot}$, for the case 1, we compare Eq.~\ref{eq:needed} and Eq.~\ref{eq:case1}. 
It is clear that only a few nova explosions are enough to cause the required amount of Li accretion.
While this seems very promising, this scenario has some difficulties in explaining the observed properties of Li-rich stars.
As previously discussed, this scenario presents difficulties when confronted with the lack of close binary companions for most unevolved Li-rich stars. 
The implied very-low-mass end states ($<0.5\,\mathrm{M_\odot}$) of this channel \citep{Kemp2021a} are also inconsistent with the effective temperatures and surface gravities of the Li-rich stars.

\begin{figure}
\includegraphics[width=1.0\hsize]{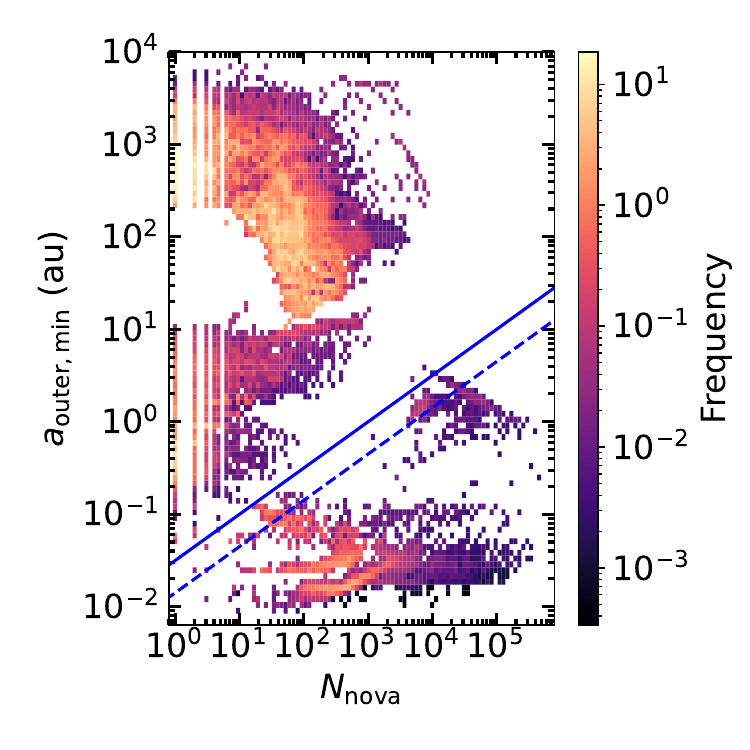}
\caption{The distribution of nova systems in the space of the minimum separation between the nova and the tertiary ($a_{\rm outer, min}$) for the stability of the triple-star system and the number of nova explosions ($N_{\rm nova}$). The blue dashed line corresponds to the condition Eq.~\ref{eq:condition}, while the blue solid line marks the selection of promising models. The reason for the use of the solid line instead of the dashed line is explained in the text.
\label{fig:minsep_Nnova}}
\end{figure}

\begin{figure}
    \includegraphics[width=1.0\hsize]{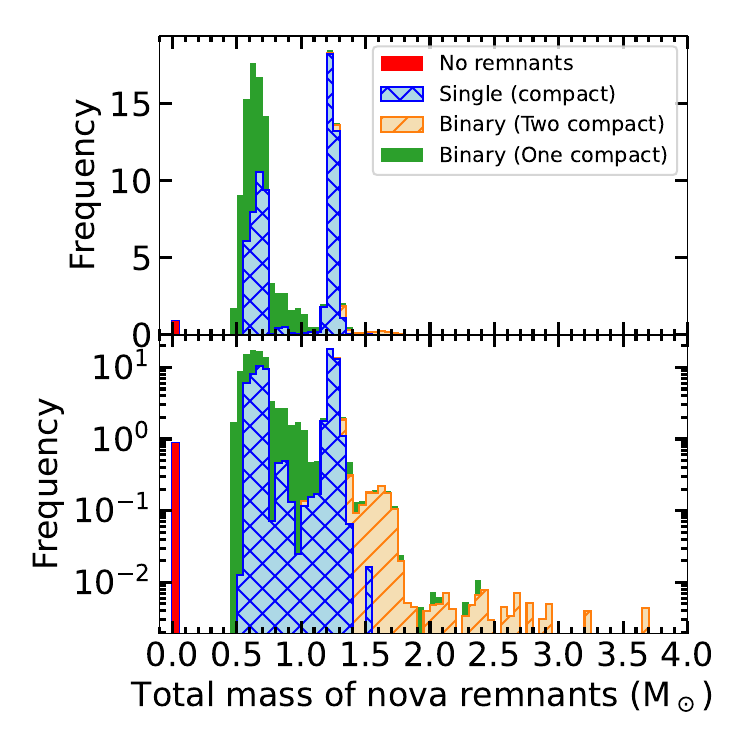}
\caption{A stacked histogram showing the total mass of remnant systems. The systems are classified by the number of remnants and the number of compact remnants. The upper panel is with the linear scale on the $y$-axis and the lower panel is with the log scale. \label{fig:end_state}}
\end{figure}

We therefore discard case 1, and proceed to assess the possibility of tertiary pollution described in case 2.
In this scenario, the separation between the nova and the tertiary has to be wide enough for the system to be dynamically stable and close enough for the tertiary to be sufficiently polluted by the nova ejecta.
We consider triple-star systems consisting of an inner binary that is drawn from the population synthesis model of \citet{Kemp2021a} and an outer tertiary with a mass of $0.8\,\mathrm{M_\odot}$.
We estimate the minimum separation between the nova and the tertiary over the period between the first and the last H nova explosion of the inner binary system using the stability criterion of \citet{Eggleton1995a}.
The largest value of this period is taken as the minimum separation for the tertiary to be on a stable orbit ($a_{\rm outer, min}$)\footnote{We also made sure that neither of the inner binary components has a radius exceeding $a_{\rm outer, min}$.}.
We then discuss if the condition Eq. \ref{eq:condition} can be met for the tertiary with the separation of $a_{\rm outer, min}$. 

Since the parameters that can vary over the widest parameter ranges among different systems in Eq. \ref{eq:needed} and \ref{eq:case2} are $N_{\rm nova}$ and $d$, the latter of which is now set to $a_{\rm outer, min}$, we show the number of nova explosions ($N_{\rm nova}$) and $a_{\rm outer, min}$ in \fig{fig:minsep_Nnova}. 
With all the parameters fixed to the default, the blue dashed line corresponds to the condition shown in Eq.~\ref{eq:condition} and the solid line running parallel to it corresponds to $M_{\rm env}=5M_{\rm acc}({\rm Li})$.
The latter runs through a region where the number of systems is significantly reduced, which allows us to more clearly separate the systems into two groups by the solid line than by the dashed line. 
Motivated by the clear separation by the solid line and its proximity to the condition Eq.~\ref{eq:condition}, we select models below this line as those promising for sufficient Li accretion from the inner binary to the tertiary.
As the parameters such as $X_{\rm Li}$ and $M_{\rm ej}$ still contain large uncertainties, we consider the use of the solid line instead of the dashed line as a reasonable choice.

This nova pollution scenario predicts the presence of Li-rich stars without a companion and those with massive compact companions.
Figure~\ref{fig:end_state} shows the end states of the selected promissing binary systems.
Interestingly, there is a peak at around the total mass of $1.25\,\mathrm{M_\odot}$, which is close to the companion masses of \nscanone\ and \nscanfive and the white dwarf mass of the most favored models in the abundance comparisons presented in \fig{fig:pattern_summary}. 
There is a tail toward the high mass end among the systems with two compact remnants, which might be an explanation for the companion of \gaians.
It is also promising that some, although not many, nova systems leave no remnants; their tertiary would be observed as a single star like most Li-rich stars.

We should, however, note that the nova population synthesis model of \citet{Kemp2021a} is based on evolutions in binary systems. 
Ideally, a dedicated population synthesis model of triple star models is needed for further investigation.
First, it needs to be investigated how the outer tertiary can be brought onto an orbit with a sufficiently small separation before the inner binary starts its nova phase.
Binary systems going through nova explosions often experience orbital shrinking due to common envelope evolutions. 
If we were to make a figure similar to \fig{fig:minsep_Nnova} using the minimum separation of a stable triple system at the initial condition of the binary system, almost no models would be selected as promising.
Thus, while there are stable orbits for the tertiary to be Li-rich during the nova explosions, the same orbits are not stable when stars are born.

Second, the evolution of triple star systems needs to be followed after the last nova explosion in order to make a comparison with the binary status of the Li-rich stars.
The single compact remnant cases in \fig{fig:end_state} form through a merger of two objects after the last nova explosion in the system.
The system often loses significant mass during this merger process.
This raises the question of whether the tertiary can avoid further mass accretion on top of the Li-rich surface and if the tertiary and the remnant can remain bound.
This point is a more serious concern for the \citetalias{El-Badry2024a} systems, as their companions are found at the distance of $1-3$ au. 
Due to the mass loss, the separation between the tertiary and the single compact remnant will be widened, and thus it might be difficult to reproduce the binary separation and the mass of the unseen companion of the \citetalias{El-Badry2024a} systems.
The modelling of triple-star systems by \citet{Shariat2023a,Shariat2025a} confirm this finding; the minimum separation possible seems to be in the range of $5-10$ au. 
Future consistent modelling of nova explosions in triple-star systems is desired to conclude if the formation of Li-rich stars with massive compact companions is possible in the nova pollution scenario.
 
The presence of the tertiary will likely change the evolution of the inner binary system as well.
It might reduce the orbital separation of the inner binary and hence enhance the interaction through the Kozai-Lidov mechanism or dynamical instability, which might lead to a shorter recurrence time of novae \citep{Knigge2022a} and/or the enhanced formation rate of nova systems ending up as type~Ia supernova and leaving no remnants \citep{Rajamuthukumar2023a}.

\subsection{Association with the ED-3 stream}

An interesting property of \objname\ is that it was identified as a member of ED-3. 
If the progenitor of the stream is a globular cluster, the enhanced N and Na abundances might result from the same formation mechanism as the second-generation stars in globular clusters \citep[see ][for a review]{Bastian2018a}.
However, the metallicity of \objname\ is significantly lower than the majority of the other stars in the stream by $\sim 0.4$ dex \citepalias{Dodd2024a}, indicating that the star does not originate from the same gas cloud as the other stars in the stream if the main progenitor of the stream is a disrupted globular cluster.
Since it is still possible that \objname\ originates from a different globular cluster, future O abundance measurements are of interest to see if the star is a second-generation star from a globular cluster. 

If \objname\ is confirmed to have a globular cluster origin, it will join NGC~6397 \#1657 studied by \citet{Koch2011a} as an unevolved Li-rich star in globular clusters.
The size of such a sample is still small, but it is interesting to see if the formation of unevolved Li-rich stars and multiple populations in globular clusters are related. 
We, however, note that no globular cluster is known to host a Li-rich population, and thus, no clear evidence for the preferred formation of Li-rich stars in globular clusters is known so far.
Furthermore, NGC~6397 \#1657 does not show N or Na enhancements \citep{Pasquini2014a}, features observed in \objname. 
This difference implies that there is likely no relation between the formation of unevolved Li-rich stars and the formation of multiple populations in globular clusters.

An alternative scenario for the formation of the ED-3 stream is that it was a small dwarf galaxy that includes a globular cluster \citepalias{Dodd2024a}.
In this case, it is possible that \objname\ is a field star in the dwarf galaxy. 
It is not clear if the formation environment plays a significant role in the formation of unevolved Li-rich stars. 
In order to investigate this with a larger sample, it is necessary to search for unevolved Li-rich stars among stars that have formed in dwarf galaxies and now in the Milky Way halo since unevolved stars in surviving dwarf galaxies are out of reach of high-resolution spectroscopy.
Such accreted stars can be identified through their kinematics and chemical abundances \citep[e.g.,][]{Nissen2010a,Helmi2018a}

\section{Conclusion\label{sec:conclusion}}

In this paper, we report on the discovery of one new unevolved Li-rich star with a low-metallicity ($\feh=-2.1$) and the detailed abundance patterns for three similarly Li-rich stars in \citetalias{El-Badry2024a}.
While \objname\ seems to be a single star, the latter three stars, \nscanone, \nscanfive, and \gaians, are known to host a relatively massive non-luminous companion.
\objname\ and two of the three stars in \citetalias{El-Badry2024a} show Na enhancements in addition to their Li-excess, a common feature in Li-rich stars.
\objname\ also shows a significant N-excess and \gaians's N abundance might also be enhanced, but N measurements for the other objects are difficult due to the wavelength coverage. 
No noteworthy peculiarities are seen in other elements, including C, Al, and $s$-process elements for the four objects.
Based on the similarities in abundance ratios, we suggest that the unevolved Li-rich stars at low-metallicity and the metal-poor stars in \citetalias{El-Badry2024a} are in the same class of objects.
This would then require formation models of unevolved Li-rich stars to reproduce Li-rich stars both with and without close companions.

We also discuss the possible polluters that provided the excess Li, N, and Na to the unevolved Li-rich stars from the abundance pattern of \objname, covering C, N, Na, Mg, Al, and Si.
The considered polluters include AGB stars modelled by \citet{Karakas2010a}, \citet{Ventura2013a}, \citet{Cristallo2015a}, \citet{Ritter2018a}, and \citet{Doherty2014a} and novae modelled by \citet{Jose1998a}, \citet{Jose2020a}, \citet{Starrfield2020a}, and \citet{Starrfield2024a}.
Among the AGB models, while two AGB models with $3$ and $3.5\,\mathrm{M_\odot}$ from \citet{Ventura2013a} show the best matches, the result is clearly highly model dependent since AGB yields calculated by different groups show a wide variety. 
In case of pollution from a nova, nova models involving massive ONe white dwarfs show as good fits as most AGB models do.
In either case, further theoretical refinements and observational constraints are needed to make a more robust conclusion; we need more evidence to select the most realistic AGB model, and we need nova models calculated at low metallicity as well as their tests against observations. 


While both the AGB and nova scenarios can reproduce the observed abundance pattern of \objname, the AGB scenarios predict the presence of a white dwarf companion, which is inconsistent with the observed binary frequency among unevolved Li-rich stars.
This inconsistency motivated us to further investigate the nova scenario.
We studied the potential configurations that can lead to sufficient mass accretion to the currently observed star in the nova scenario.
For unevolved Li-rich stars, the only viable configuration is in triple star systems in which the currently observed star is the outer tertiary to the inner binary which forms a nova system. 
We select promising binary configurations that can host a tertiary on a dynamically stable orbit with a small separation and lead to a large number of nova explosions so that sufficient pollution can happen. 
Interestingly, some, although not many, of such binary systems leave no remnants, which is consistent with the observations that show a significant fraction of unevolved Li-rich stars does not show any radial velocity variation.
They could also leave remnants, a single compact object, binary compact objects, or a binary consisting of a compact object and a star, which could explain the presence of Li-rich stars with massive non-luminous companions, such as those from \citetalias{El-Badry2024a}.

The properties of the Li-enhanced binaries discovered by \citetalias{El-Badry2024a} are most readily explained in the nova scenario if the luminous stars are tertiaries orbiting inner binaries containing a massive ONe white dwarf accreting from a low-mass star or brown dwarf. Given the systems' ancient ages, it is plausible that the accretion rates in the inner binaries would have fallen to sufficiently low levels that no signatures of ongoing accretion are detectable \citep{Knigge2011a}. Whether such a scenario can be realized with realistic triple evolution models remains to be seen.

While this paper presents a proof of concept to the nova-pollution scenario, further observational and theoretical studies are needed for a more robust conclusion. 
First, more intensive radial velocity monitoring is needed to quantify the binary frequency among unevolved Li-rich stars.
Second, measurements of detailed abundance patterns, including C, N, O, Na, Mg, and Al, for more objects are needed to obtain better constraints on the polluters. 
In particular, it is of interest to study if the N enhancements seen in \objname\ is a common feature among such stars.
Theoretically, a population synthesis model of triple star models is necessary to fully explore the viability of the nova pollution scenario. 
Population synthesis and nucleosynthesis models also need to be studied at low-metallicity.

A measurement of the N isotope ratio provides a robust test for separating the nova and AGB pollution scenarios.
Both nova and AGB models can produce enhanced N, but the N isotope ratio is expected to be different between the two sources.  
If the enhanced N mainly comes from nova, some stars might have very high $^{15}$N abundance, and the isotope ratio $\mathrm{^{15}N/^{14}N}$ could exceed 1 \citep{Jose1998a}.
On the other hand, AGB models are not expected to produce a detectable amount of $^{15}$N.
We thus encourage future attempts to measure the N isotope ratio from analysis of molecular lines in very high resolution, high $S/N$ spectra of unevolved Li-rich stars.

\begin{acknowledgements}
    We thank Ko Takahashi and Takuma Suda for the helpful discussions. 
    We also thank Luca Sbordone for his advice on the analysis of the H$\alpha$ line.
    TM is supported by a Gliese Fellowship at the Zentrum f\"ur Astronomie, University of Heidelberg, Germany. This research was supported in part by Grants-in-Aid for Scientific Research (24K07040) from Japan Society for the Promotion of Science and by a Spinoza Grant from the Dutch Research Council (NWO) awarded to AH.

\end{acknowledgements}
    
\bibliographystyle{aa_url}
\bibliography{ref.bib}

\appendix
\section{Spectrum of \objname\label{app:spectrum}}
\begin{figure*}
    \begin{minipage}{0.5\hsize}
     \includegraphics[width=\hsize]{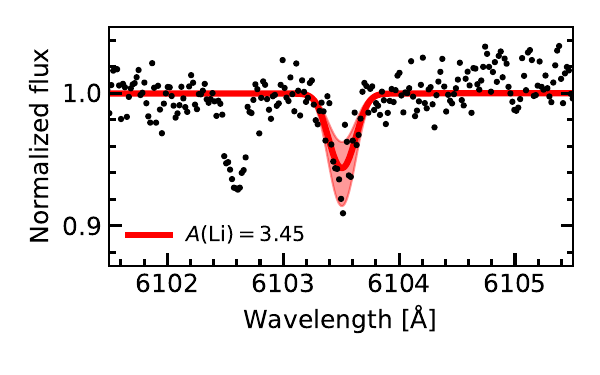}
    \end{minipage}   
    \begin{minipage}{0.5\hsize}
     \includegraphics[width=\hsize]{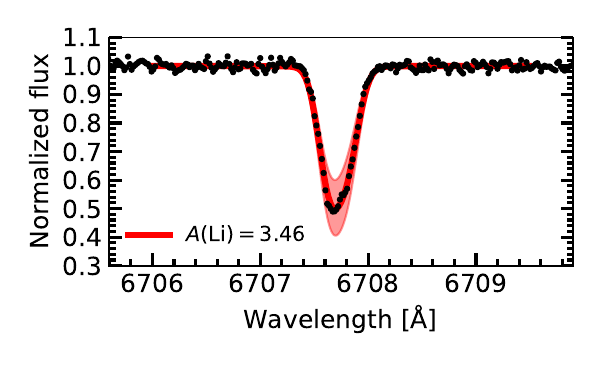}
    \end{minipage}\vspace{-5mm}
    \begin{minipage}{1.0\hsize}
        \includegraphics[width=\hsize]{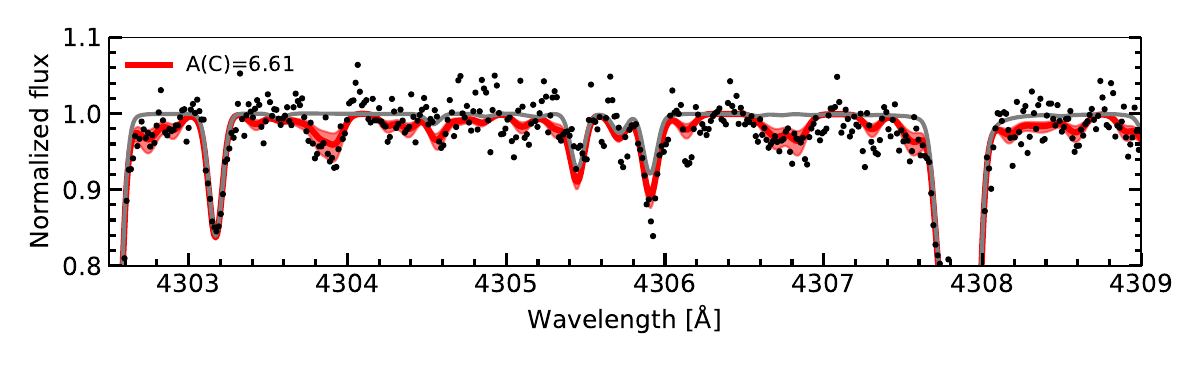}
    \end{minipage}\vspace{-5mm}
    \begin{minipage}{1.0\hsize}
        \includegraphics[width=\hsize]{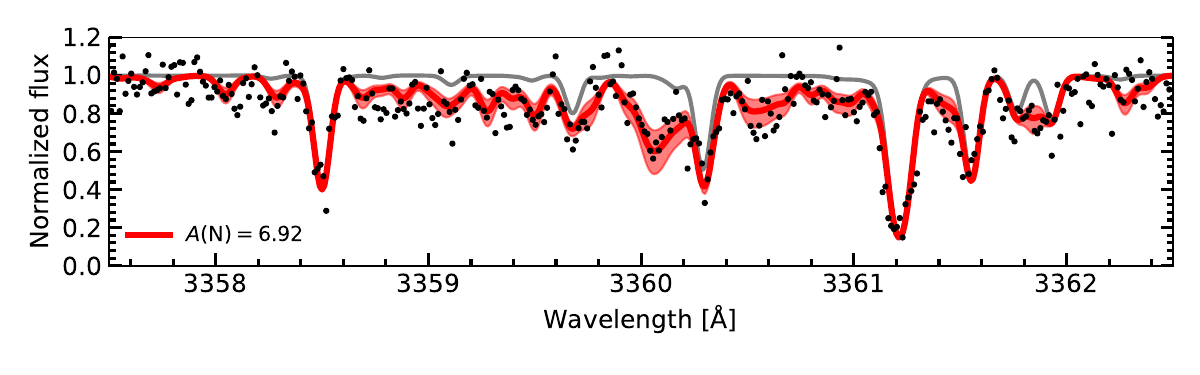}
    \end{minipage}\vspace{-5mm}
    \begin{minipage}{0.5\hsize}
        \includegraphics[width=\hsize]{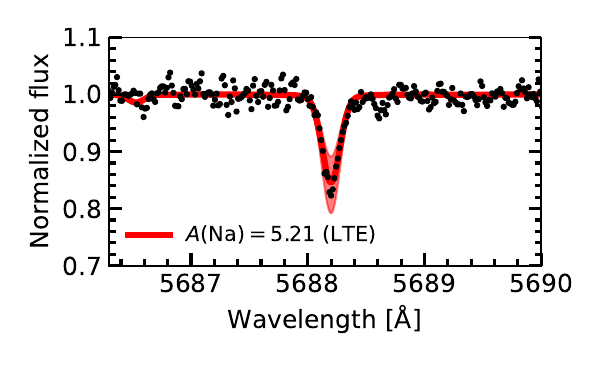}
    \end{minipage}
    \begin{minipage}{0.5\hsize}
        \includegraphics[width=\hsize]{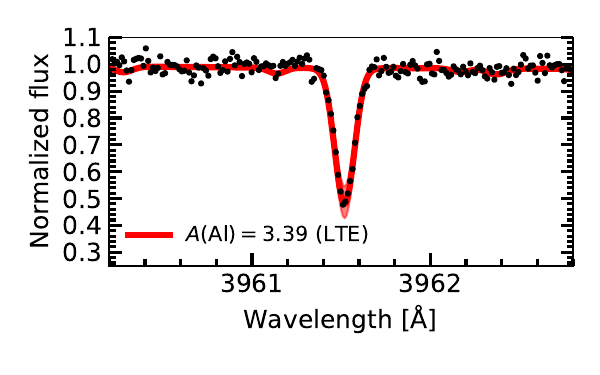}
    \end{minipage}
    \caption{Portion of the spectrum of \objname. The observed spectrum is shown with black dots, and the best-fit synthetic spectra are shown with red lines. The shaded regions show spectra computed with the abundance of the element of $\pm 0.2$ dex from the best-fit. The gray lines in the middle two panels show the synthetic spectra with extremely low abundances of C and N, respectively. Note that while spectral fitting was done in LTE, non-LTE correction was applied to Na and Al abundances. \label{fig:spectra}}
\end{figure*}

\end{document}